\documentclass{appolb}
\usepackage{graphicx}

\usepackage{hyperref}
\usepackage{lineno}
\usepackage{color}
\usepackage{caption}
\usepackage{multirow}

\usepackage{graphics}
\usepackage{epstopdf}
\usepackage{amssymb}
\usepackage{xspace}
\usepackage{amsmath}
\usepackage{upgreek}
\usepackage{hyperref}

\newcommand{\snn}{$\sqrt{s_{\mathrm{NN}}}~$}

\newcommand{\pt}{$p_{\mathrm{T}}~$}

\newcommand{\ptt}{\ensuremath{p_{\mathrm{T}}^{\rm trig}}\xspace}

\newcommand{\py}{PYTHIA~8\xspace}
\newcommand{\ep}{EPOS~LHC\xspace}

\newcommand{\pp}           {pp\xspace}

\newcommand{\PbPb}         {\mbox{Pb--Pb}\xspace}

\newcommand{\pPb}          {\mbox{p--Pb}\xspace}

\newcommand{\GeVc}         {Ge\kern-.1emV/$c$\xspace}
\newcommand{\MeVc}         {Me\kern-.1emV/$c$\xspace}
\newcommand{\TeV}          {Te\kern-.1emV\xspace}
\newcommand{\GeV}          {Ge\kern-.1emV\xspace}
\newcommand{\MeV}          {Me\kern-.1emV\xspace}
\newcommand{\GeVmass}      {Ge\kern-.2emV/$c^2$\xspace}
\newcommand{\MeVmass}      {Me\kern-.2emV/$c^2$\xspace}

\newcommand{\SPD}          {\rm{SPD}\xspace}

\newcommand{\VZERO}        {\rm{V0}\xspace}
\newcommand{\VZEROA}       {\rm{V0A}\xspace}
\newcommand{\VZEROC}       {\rm{V0C}\xspace}

\begin{document}
%\linenumbers

\title{Particle production as a function of underlying-event activity and search for jet-like modifications in pp, p--Pb, and Pb--Pb collisions at $\sqrt{s_{\rm NN}}=5.02$\,TeV with ALICE%
\thanks{Presented at the XXIXth International Conference on Ultra-relativistic Nucleus-Nucleus Collisions}%
}
\author{Antonio Ortiz, for the ALICE collaboration
\address{CERN, 1211 Geneva 23, Switzerland\\UNAM, Apartado Postal 70-543, Ciudad de M\'exico 04510, M\'exico}
}

\maketitle
\begin{abstract}
The similarity between small- and large-collision systems is explored using the charged-particle multiplicity in the transverse region, $N^{\rm T}_{\rm ch}$, which is sensitive to the underlying event.  Measurements of charged-particle production as a function of  $N^{\rm T}_{\rm ch}$  in pp, p--Pb and Pb--Pb collisions at $\sqrt{s_{\rm NN}}=5.02$ TeV in the toward, away and transverse regions are discussed. These three regions are defined relative to the track with the largest transverse momentum in the event  ($p_{\rm T}^{\rm trig}$). The activity in the transverse region is subtracted from the activity in the toward and the away regions to search for jet-like modifications in small-collision systems. The jet-like signals are studied both as a function of  $N^{\rm T}_{\rm ch}$  and $p_{\rm T}^{\rm trig}$.  Results are compared with two general purpose Monte Carlo event generators: PYTHIA~8 and EPOS~LHC. 

\end{abstract}
  
\section{Introduction}

In models incorporating multi-parton interactions (MPI), particles produced in the hard scattering (jet) are accompanied by particles from additional parton-parton interactions, as well as from the proton break-up~\cite{Sjostrand:1987su}. This component of the collision makes up the underlying event (UE). The traditional UE analysis focuses on the study of particles in three topological regions depending on their azimuthal angle relative to the leading particle ($|\Delta\varphi| =|\varphi-\varphi^{\mathrm{trig}}|$), which is the one with the highest transverse momentum in the event ($p_{\rm T}^{\rm trig}$). The toward region ($|\Delta\varphi|<\pi/3$\,rad) contains the primary jet and UE, while the away region ($\pi/3\,{\rm rad} < |\Delta\varphi| < 2\pi/3\, {\rm rad}$) contains the fragments of the recoil jet and UE. In contrast, the transverse region ($|\Delta\varphi|>2\pi/3$\,rad) is dominated by the UE dynamics, but it also includes contributions from initial- and final-state radiation~\cite{Bencedi:2021tst}.  
%$\mbox{ \pi/3\,{\rm rad} < |\Delta\varphi| < 2\pi/3\, {\rm rad} }$
High-energy \pp and \pPb collisions (small-collision systems) unveiled remarkable similarities with heavy-ion collisions like collectivity, and strangeness enhancement. In heavy-ion collisions such effects are attributed to the formation of the strongly-interacting quark-gluon plasma (sQGP)~\cite{Busza:2018rrf}. Whether or not a small drop of sQGP is formed in \pp and \pPb collisions, is still a matter of debate, in particular, because jet quenching signals have not been observed so far in small-collision systems. This work discusses new results aiming at understanding small-collision systems using UE-inspired techniques~\cite{ALICE:2022fnb,ALICE:2022qxg}, with special emphasis on quantities which by construction are expected to be sensitive to jet quenching. 

In the first part of the analysis, the average charged-particle multiplicity density (number density) in each topological region are studied as a function of the transverse momentum of the leading particle. Measurements from \pp and \pPb collisions at $\sqrt{s_{\rm NN}}=5.02$\,TeV are reported. In the second part of the analysis, the high-\pt particle production in \pp, \pPb and \PbPb collisions at $\sqrt{s_{\rm NN}}=5.02$\,TeV are measured as a function of the average multiplicity in the transverse region ($N^{\rm T}_{\rm ch}$). In this way the jet contribution is explicitly excluded from the multiplicity estimator, thus reducing the selection biases~\cite{Martin:2016igp,Weber:2018ddv}. The high-\pt yield in the toward and away regions after subtracting the underlying-event contribution are reported for different multiplicity classes. The multiplicity dependent yields in the jet-like signals are compared with analogous measurements in minimum-bias (MB) \pp collisions. This allows to search for modifications of particle production in the jet-like signals in events with large underlying-event activity. 

\section{Experimental setup and data analysis}

This analysis is based on data recorded by ALICE during the \pp, \pPb, and  \PbPb runs at \snn = 5.02\,\TeV. The \VZERO detector, and the Silicon Pixel Detector (\SPD) are used for triggering and background rejection. The \VZERO consists of two arrays of scintillating tiles placed on each side of the interaction point covering the pseudorapidity intervals of 2.8~$<\eta<$~5.1 (\VZEROA) and -3.7~$<\eta<$~-1.7 (\VZEROC).  The data were collected using a MB trigger, which required a signal in both \VZEROA and \VZEROC detectors. A criterion based on the offline reconstruction of multiple primary vertices in the SPD is applied to eliminate the pileup contamination.  For the multiplicity dependent studies in \pp and \PbPb collisions, the sample is subdivided into different multiplicity classes based on the total charge deposited in both \VZERO sub-detectors (V0M amplitude). For \pPb collisions, the sample is subdivided based on the total charge deposited in the \VZEROA sub-detector (\VZEROA amplitude).  The \VZEROA amplitude selection has a small multiplicity fluctuation bias due to the enhanced contribution from the Pb-fragmentation region.

The \pt spectra are measured in the central barrel with the Inner Tracking System and the Time Projection Chamber following the standard procedure of the ALICE collaboration~\cite{Acharya:2018qsh}. Tracks are selected in the kinematic ranges $p_{\rm T}>0.5$\,GeV/$c$ and $|\eta|<0.8$. The raw yields as a function of \ptt or the activity in the \VZERO detector are corrected for efficiency and contamination from secondary particles. The efficiency correction is calculated from Monte Carlo simulations with GEANT3~\cite{Brun:1082634} transport code, which made use of PYTHIA~8 (Monash)~\cite{Skands:2014pea}, EPOS-LHC~\cite{Pierog:2013ria} and HIJING~\cite{Deng:2010mv} event generators for \pp, \pPb and \PbPb collisions, respectively.  Since the event generators do not reproduce the relative abundances of different particle species in the real data, the efficiency obtained from the simulations is re-weighted considering the particle composition from data~\cite{Acharya:2018qsh}. A multi-component template fit based on the distance-of-closest-approach distributions from the simulation is used for the estimation of secondary contamination~\cite{Acharya:2018qsh}. Another correction which is relevant for \ptt$<5$\,GeV/$c$ is the leading track misidentification correction. In short, the leading particle may not be detected due to finite acceptance and efficiency of the detection apparatus, and a lower \pt track enters the analysis instead. If the misidentified leading particle has a different \pt but roughly the same direction as the true leading particle, this leads to a shift in \ptt. On the other hand, if the misidentified leading particle has a significantly different direction than the true one, this will cause a rotation of the event topology and a bias on the UE observables. The data are corrected for these effects using a data-driven procedure~\cite{ALICE:2022fnb}.

The main sources of systematic uncertainties include: 1) track selection criteria,  2) imperfect simulation of the detector response, 3) secondary particle contamination, 4) leading track misidentification, and 5) correction method (Monte Carlo non closure). For the estimation of the total systematic uncertainty, all the contributions were summed in quadrature.  

\section{Results and discussion}

\begin{figure}[tb]
\begin{center}
    \includegraphics[width=7.70cm]{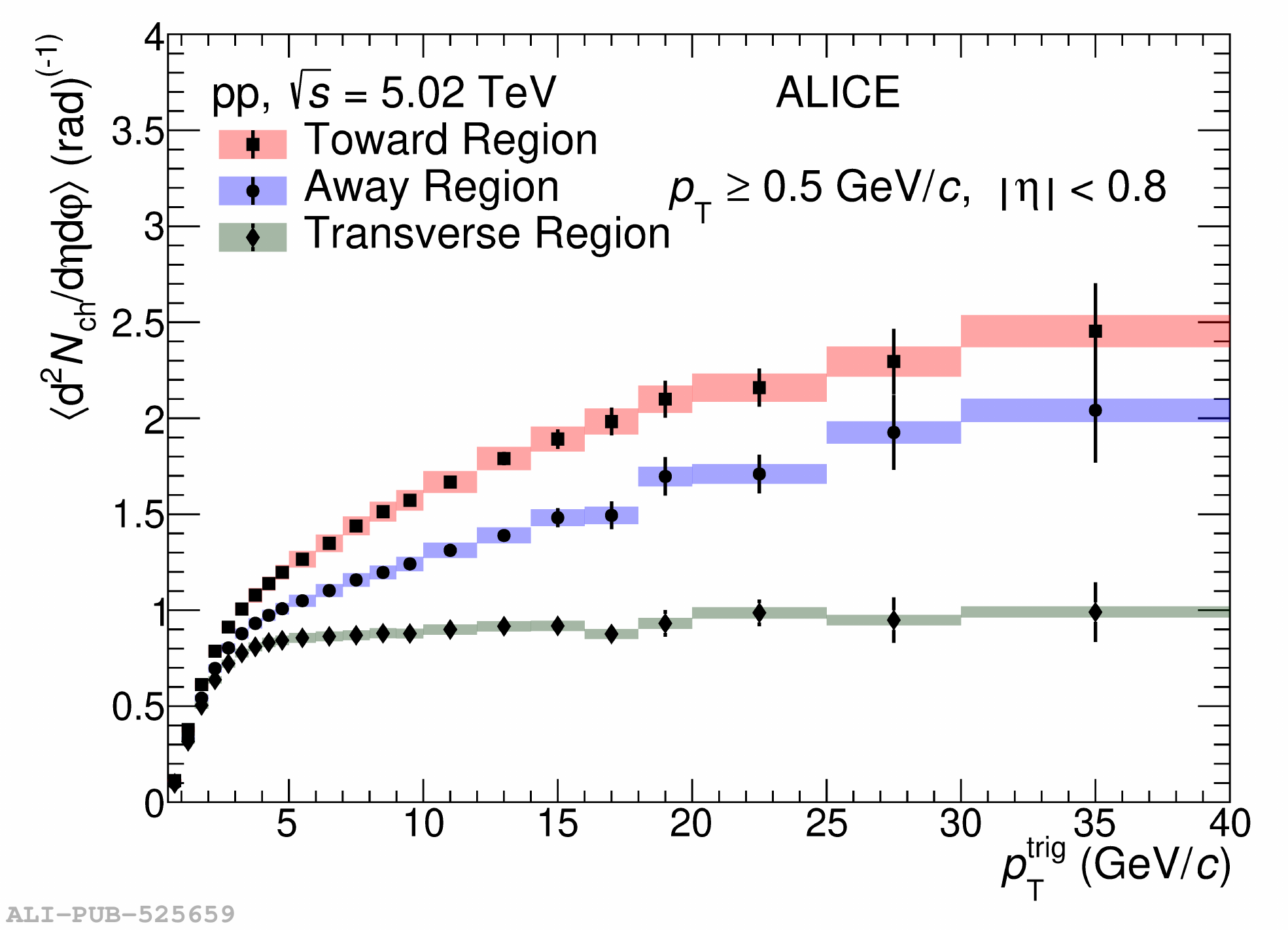}
    \includegraphics[width=7.70cm]{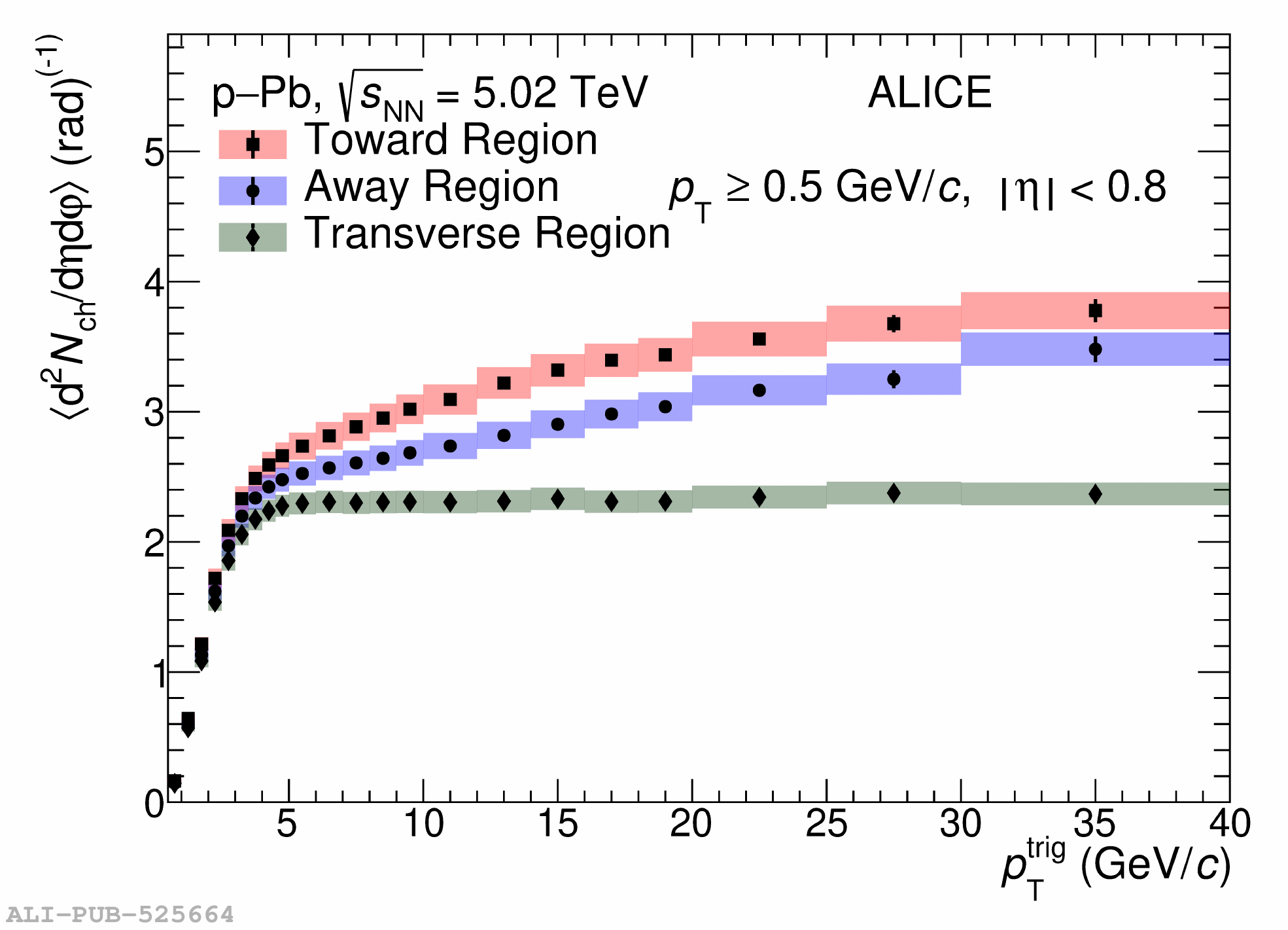}
\end{center}
    \caption{Number density as a function of \ptt measured in \pp (left) and \pPb collisions (right) at $\sqrt{s_{\rm NN}}=5.02$\,TeV. Results for the toward, transverse, and away regions are displayed. The boxes and the error bars represent the systematic and statistical uncertainties, respectively.}
    \label{fig:10011}
\end{figure}

Figure~\ref{fig:10011} shows the number density as a function of \ptt measured in \pp and \pPb collisions. Results are presented for the toward, transverse, and away regions. The $p_{\rm T }^{\rm trig}$ dependence in all regions is similar for both collision systems. For \ptt below 5\,GeV/$c$ the number densities exhibit a steep rise with increasing \ptt. For higher \ptt the increases are less steep. The effect is more pronounced for the transverse region, where the number density is almost independent of \ptt. This saturation is expected in models that include the concept of impact parameter such that the requirement of the presence of a high-\pt particle in a \pp collision biases the selection of collisions towards those with a small impact parameter~\cite{Strikman:2011cx}. The remarkable similarity between \pp and \pPb collisions indicates a similar bias in the \pPb impact parameter, and to some extend, in the nucleon-nucleon impact parameter~\cite{ALICE:2014xsp}. On the other hand, the rise of the number density observed for the toward and away regions is due to the contribution from the jet fragments. However, for these two topological regions the number densities increase faster with \ptt in \pp collisions than in \pPb collisions. This effect is due to a larger contribution from UE to the toward and away regions in \pPb collisions compared to \pp collisions, which is expected given that in \pPb collisions more than one nucleon-nucleon interaction can occur.

\begin{figure}[!tb]
 \begin{center}
 \includegraphics[width=7.70cm]{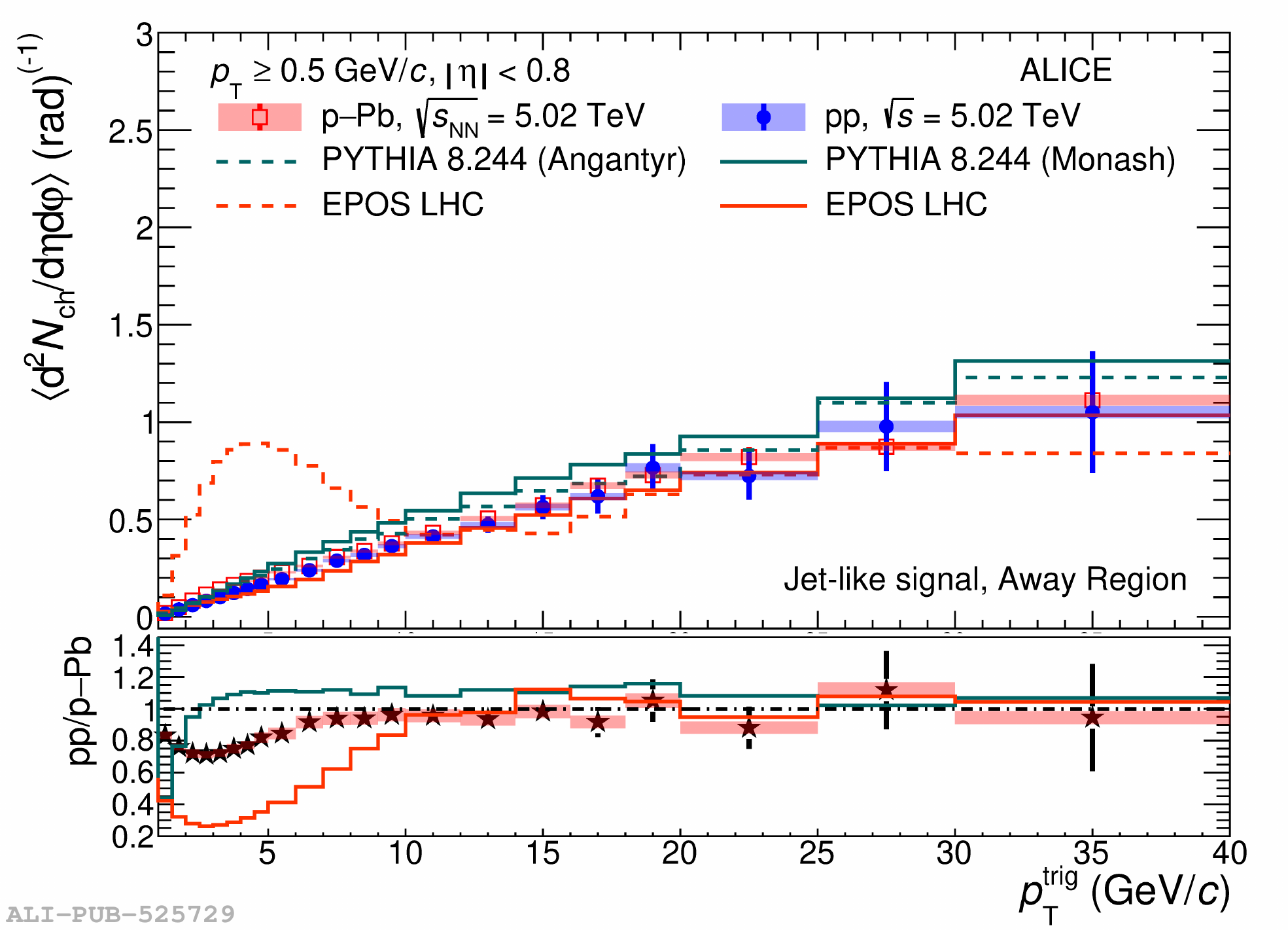}
 \includegraphics[width=7.70cm]{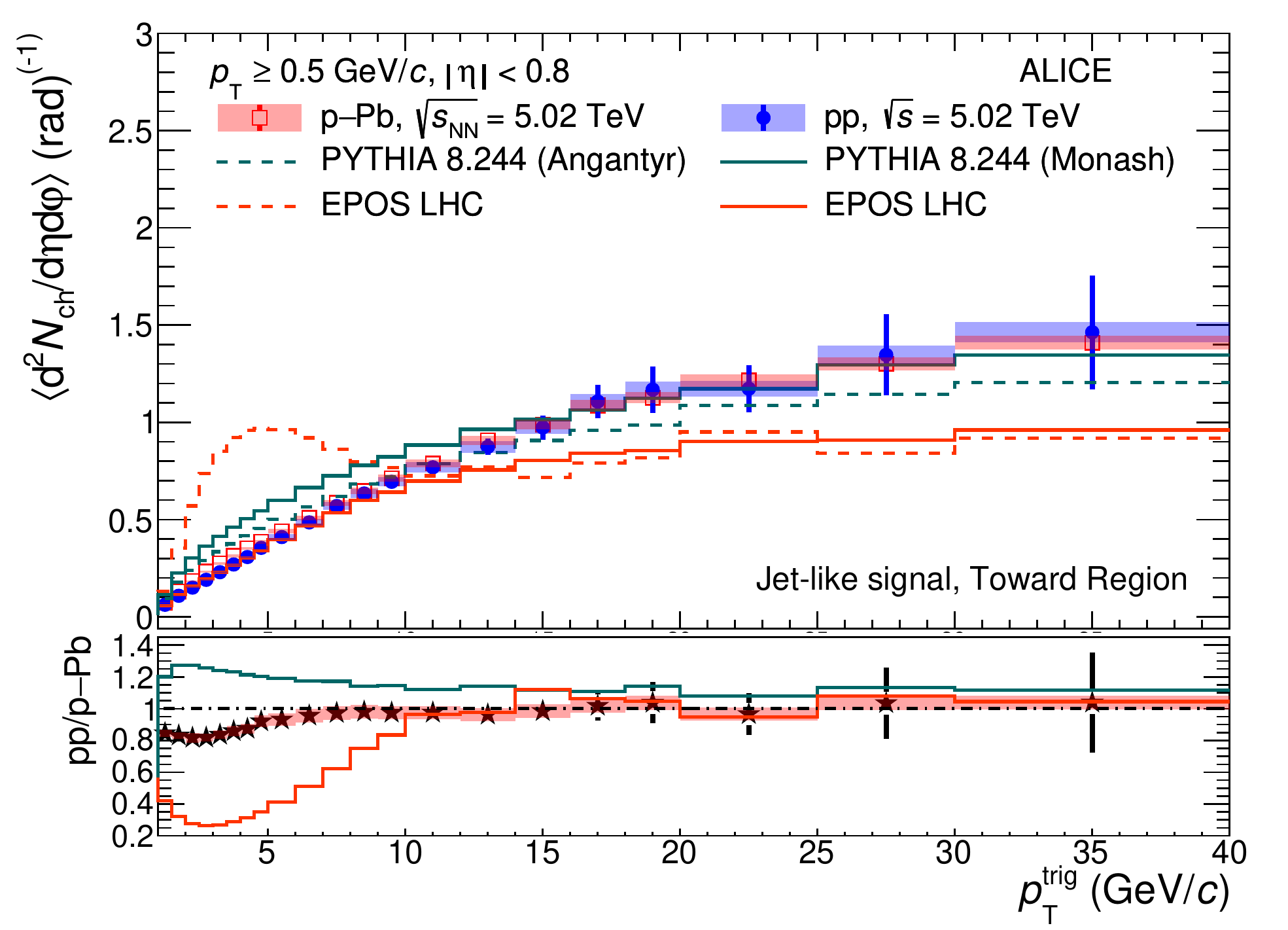}  
 \caption{Number density as a function of \ptt in \pp and \pPb collisions at $\sqrt{s_{\rm NN}}=5.02$\,\TeV. Results for the jet-like signals in the away (left) and toward (right) regions are shown. The \pp results scaled to the \pPb ones are displayed in the bottom panel.}
 \label{fig:f57212}
 \end{center}  
 \end{figure}

As discussed earlier, requiring a high-\pt particle makes an indirect selection of collisions with small impact parameters possible. Thus, it is interesting to compare the particle produced in the hard scattering in \pp and \pPb collisions for similar \ptt intervals.  In this way, given that the UE activity is 2.4 times larger in \pPb than in \pp collisions, one can search for possible jet-like modifications in \pPb collisions. In order to increase the sensitivity to the particles produced in the hard scattering, jet-like signals are obtained from the number density in the toward and away regions after subtracting the number density in the transverse region.  Figure~\ref{fig:f57212} shows the jet-like contribution to the number density in the toward and away regions as a function of \ptt for \pp and \pPb collisions at $\sqrt{s_{\rm NN}}=5.02$\,\TeV. The number densities in the jet-like signals rise with increasing \ptt in the entire range of the measurement.  At high \ptt ($>10$\,\GeVc), the number density exhibits a remarkable similarity between \pp and \pPb collisions. Within 10\%, both \py~Angantyr~\cite{Bierlich:2018xfw} and \ep reproduce this feature.  This is consistent with the absence of medium effects in minimum-bias \pPb collisions at high \ptt. At lower \ptt, the models overestimate the number density in \pPb collisions. The disagreement is more remarkable for \ep than for \py~Angantyr. The number density in \pp collisions scaled to that in \pPb collisions is smaller than unity, reaching a minimum of $\approx 0.8$ at $\ptt \approx 3$\,\GeVc. This behaviour is not reproduced by \py~Angantyr. In contrast, \ep exhibits a similar pattern, but the size of the effect is much larger than in data. The main difference between \py~Angantyr and \ep is that \ep incorporates collective flow, which is expected to give an effect in the \ptt interval where the differences between measurements in \pp and \pPb collisions are observed. Moreover, the Angantyr model in \textsc{PYTHIA}~8 extrapolates the dynamics from \pp collisions to \pPb and \PbPb collisions, generalising the formalism adopted for \pp collisions by including a description of the nucleon positions within the colliding nuclei and utilising the Glauber model to calculate the number of interacting nucleons and binary nucleon-nucleon collisions.

To further investigate the possible modification of the particles produced in the hard scattering using \pp and \pPb data, only collisions containing a leading track transverse momentum within $8 < p_{\rm T}^{\rm trig} < 15\,{\rm GeV}/c$ are considered. Moreover, small-collision systems are compared with \PbPb collisions at the same centre-of-mass energy per nucleon pair, where there is enough evidence of medium jet modifications due to the presence of sQGP.  The high-\pt yields ($4<p_{\rm T}<6$\,GeV/$c$) in the toward ($Y^{\rm t}$) and away ($Y^{\rm a}$) regions obtained after the subtraction of the high-\pt yield in the transverse region ($Y^{\rm T}$) are measured. The subtracted yields ($Y^{\rm{st,sa}}$) are further normalised to those measured in MB \pp collisions. With these quantities the $I_{X}^{\rm{t,a}}$ factors are defined as follows: 
\begin{equation}
I_{X}^{\rm{t,a}} \equiv
(Y_{X}^{\rm{t,a}} - Y_{X}^{\rm T}) / (Y_{\rm {pp,MB}}^{\rm{t,a}} - Y_{\rm {pp,MB}}^{\rm T})=Y_{X}^{\rm{st,sa}} / Y_{\rm {pp,MB}}^{\rm{st,sa}},
\end{equation} 
where $X$ indicates the collision system and the event multiplicity class. In the absence of medium effects or selection biases, this quantity is expected to be consistent with unity.

\begin{figure}[ht!]
\centering
\includegraphics[width=36pc]{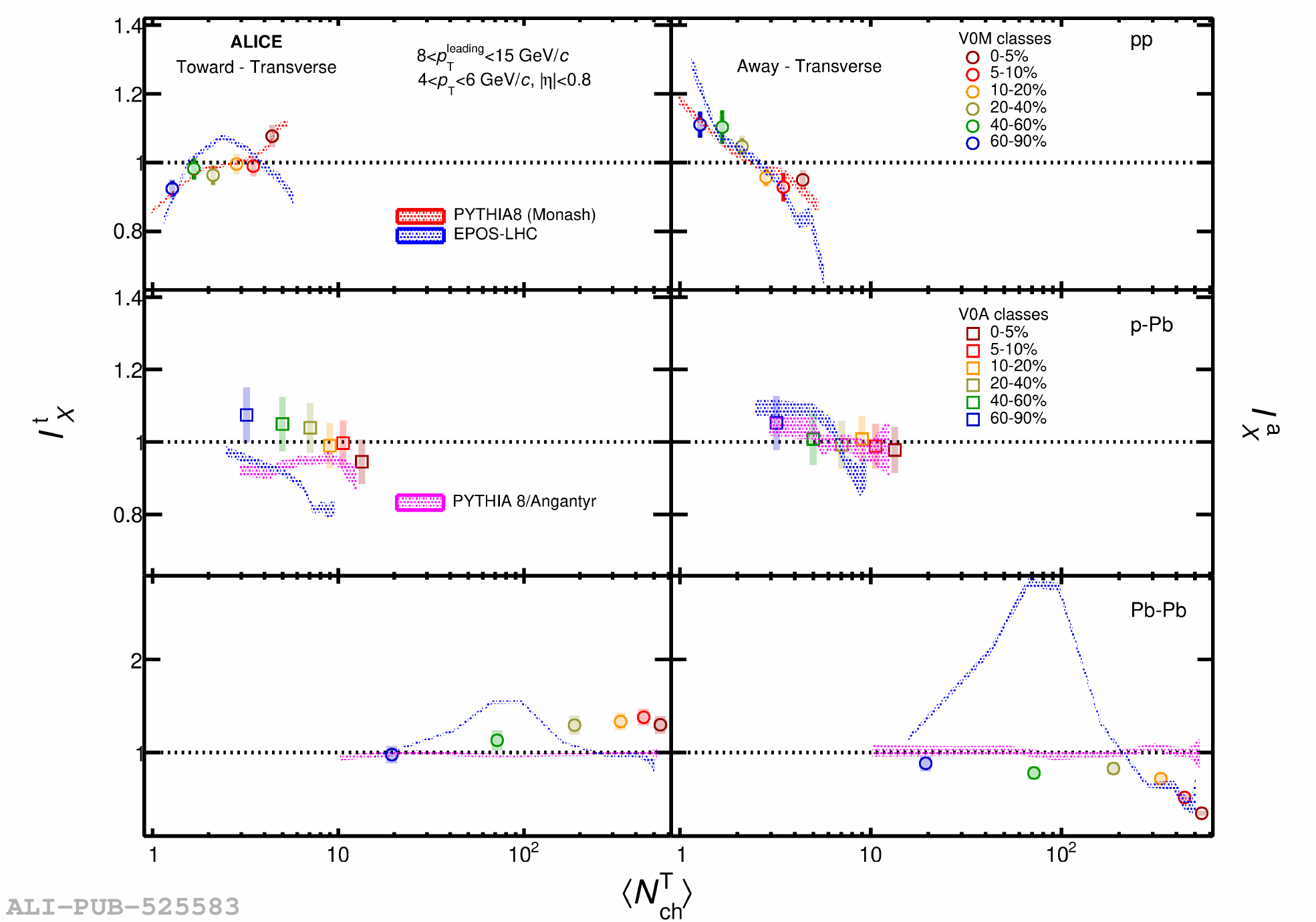}
\caption{Comparison of the measured $I_{X}^{\rm t}$ (left) and $I_{X}^{\rm a}$ (right) in 4 $< p_{\rm T} < $ 6 \GeVc with model predictions. The results are shown as a function of $\langle N_{\rm ch}^{\rm T} \rangle$ for different multiplicity classes in \pp (top panel), \pPb (middle panel) and \PbPb (bottom panel) collisions at \snn $=$ 5.02\,\TeV. The red and magenta lines show the \textsc{PYTHIA}~8 (Monash)~\cite{Skands:2014pea} and \textsc{PYTHIA}~8~Angantyr~\cite{Skands:2014pea} predictions, respectively. The blue lines show the EPOS-LHC~\cite{Pierog:2013ria} results. The statistical and systematic uncertainties are shown by bars and boxes, respectively.}
\label{fig3}
\end{figure}

Figure~\ref{fig3} shows $I_{\rm{X}}^{\rm{t,a}}$ as a function of $\langle N_{\rm ch}^{\rm T} \rangle$ for \pp, \pPb and \PbPb collisions. Data are compared with \py and \ep. As discussed earlier the events were classified in terms of the activity in the \VZERO detector. This reduces the presence of particles from hard Bremsstrahlung gluons that can contribute to the transverse region~\cite{Bencedi:2020qpi}. Within uncertainties, the $I_{\rm{X}}^{\rm{t,a}}$ values are close to unity for all the multiplicity classes measured in \pp collisions.  The small multiplicity dependence observed in \pp collisions is due to selection biases. Within uncertainties, the effect is well reproduced by \py, which does not incorporate any jet quenching mechanism. The origin of this small effect in high $\langle N_{\rm ch}^{\rm T} \rangle$ collisions is related to a remaining bias towards harder fragmentation and more activity from initial and final state radiation.  The results indicate that effects induced by possible energy loss in \pp collisions are not observed within uncertainties in the multiplicity and transverse momentum ranges reported in this work.

For \pPb collisions, within uncertainties, the $I_{\rm{X}}^{\rm{t,a}}$ values are close to unity for all the multiplicity classes. The data are compared to \textsc{PYTHIA}~8~Angantyr~\cite{Bierlich:2018xfw} and EPOS-LHC predictions.  The Angantyr model predicts the $I_{X}^{\rm{a}}$ factor to be consistent with unity, and $I_{X}^{\rm{t}}$ slightly below unity. On the other hand, EPOS-LHC does not describe neither the magnitude nor the trend of the multiplicity dependence of the measured ratio in the toward region, $I_{X}^{\rm{t}}$. However, the model is in reasonable agreement with data in the away region. Data from \pPb collisions indicate the absence of medium effects in both the toward and away regions within the multiplicity and transverse momentum intervals reported in this work.  

By contrast, for \PbPb collisions the $I_{\rm{X}}^{\rm{t,a}}$ values are compatible with unity for peripheral collisions, and show a gradual increase (reduction) with the increase in multiplicity for the toward (away) region. The behaviour is the same even considering the modulation of the underlying event by elliptic flow ($v_2$).  The interpretation of the results is similar to that reported in~\cite{Aamodt:2011vg}. On the one hand, the suppression in the away region is expected from the strong in-medium energy loss; on the other hand, the enhancement observed in the toward region is also subject to medium effects. The ratio is sensitive to a) a possible change of the fragmentation functions, b) a possible modification of the quark-to-gluon jet ratio in the final state due to different coupling with the medium, and c) a possible bias on the parton spectrum due to trigger particle selection. It is likely that all three effects play a role~\cite{Aamodt:2011vg}. Regarding the model comparisons, \textsc{PYTHIA}~8~Angantyr, which does not include jet quenching effects, predicts $I_{X}^{\rm{t,a}}$ values consistent with unity for all the multiplicity classes in \PbPb collisions. In EPOS-LHC, a certain \pt cutoff is defined in such a way that, above this cutoff, a particle loses part of its momentum in the core but survives as an independent particle produced by a flux tube. Soft particles, which are below the \pt cutoff, get completely absorbed and form the core. This sort of energy loss mechanism implemented in EPOS-LHC depends on the system size~\cite{Pierog:2013ria,Baier:1996kr,Peigne:2008ns}. With these implementations, EPOS-LHC predicts a significant enhancement of $I_{X}^{\rm{t,a}}$ for low $\langle N_{\rm ch}^{\rm T} \rangle$ ranges and deviates significantly from the experimental results. However, it reasonably describes $I_{X}^{\rm{a}}$ at high multiplicities.

\section{Conclusions}
For the first time the properties of the underlying event are measured in collisions involving heavy ions. The charged-particle densities as a function of \ptt exhibit a saturation at $\ptt\approx5$\,\GeVc in the transverse region for both \pp and \pPb collisions at $\sqrt{s_{\rm NN}}=5.02$\,TeV. In models like \py this effect is due to a bias towards collisions with small impact parameter. The results for the toward and away regions are also remarkably similar in \pp and \pPb collisions. In order to investigate the presence of medium effects at high \pt in jet-like signals, the charged-particle density in the toward and away regions are studied after subtracting the number density in the transverse region. For $p_{\rm T}^{\rm trig}>10$\,GeV/$c$ the number densities are compatible in \pp and \pPb collisions. An additional measurement that considers collisions containing a leading particle within $8<p_{\rm T}^{\rm trig}<15$\,\GeVc is also reported. The high-\pt yields ($4<p_{\rm T}<6$\,GeV/$c$)  are measured in different multiplicity classes for \pp, \pPb and \PbPb collisions at $\sqrt{s_{\rm NN}}=5.02$\,TeV, the yields are compared with analogous measurements in minimum-bias \pp collisions. While the data from \PbPb collisions are in agreement with expectations from parton energy loss due to the presence of a hot and dense medium, \pp and \pPb data do not show any hint of medium effects in the multiplicity and transverse momentum ranges which are reported. 

\section{Acknowledgement}
This work has been supported by CONACyT under the Grants CB No. A1-S-22917 and CF No. 2042, as well as by UNAM under the program PASPA-DGAPA.

\bibliographystyle{utphys}
\bibliography{biblio}

\end{document}